%% file: root.tex
\documentclass[letterpaper, 10 pt, conference]{ieeeconf}  

\IEEEoverridecommandlockouts                              

\usepackage{graphics} 
\usepackage{graphicx}
\usepackage{subcaption}
\usepackage{epsfig} 
\usepackage{mathptmx} 
\usepackage{times} 
\usepackage{amsmath} 
\usepackage{amssymb}  
\usepackage{algorithm}
\usepackage{algpseudocode}
\usepackage{xcolor} 
\usepackage{dsfont}
\usepackage{booktabs}
\usepackage{ulem}
\usepackage{changes}

\newcommand{\vx}{{\boldsymbol x}}

\newcommand{\calC}{\mathcal{C}} 
\newcommand{\calK}{\mathcal{K}} 

\usepackage{amsthm} 

\theoremstyle{definition}
\newtheorem{definition}{Definition}

\newtheorem{remark}{Remark}

\title{\LARGE \bf
 Time Shift Governor-Guided MPC with Collision Cone CBFs for \\Safe Adaptive Cruise Control in Dynamic Environments
}

\author{Robin Inho Kee$^{1}$, Taehyeun Kim$^{1}$, Anouck Girard and Ilya Kolmanovsky
\thanks{Robin Inho Kee is with the Department of Robotics, University of Michigan, Ann Arbor, 48109 MI, USA. {\tt\small inhokee@umich.edu}.}
\thanks{Taehyeun Kim, Anouck Girard, and Ilya Kolmanovsky are with the Department of Aerospace Engineering, University of Michigan, Ann Arbor, 48109 MI, USA. {\tt\small\{taehyeun,anouck,ilya\}@umich.edu}.}
\thanks{$^{1}$These authors contributed equally to this work.}%
\thanks{The second and fourth authors acknowledge support by the Air Force Office of Scientific Research (AFOSR) grant FA9550-23-1-0678.}
}

\begin{document}

\maketitle
\thispagestyle{empty}
\pagestyle{empty}

\begin{abstract}

This paper introduces a Time Shift Governor (TSG)-guided Model Predictive Controller with Control Barrier Functions (CBFs)-based constraints for adaptive cruise control (ACC). This MPC-CBF approach is defined for obstacle-free curved road tracking, while following distance and obstacle avoidance constraints are handled using standard CBFs and relaxed Collision Cone CBFs. In order to address scenarios involving rapidly moving obstacles or rapidly changing leading vehicle's behavior, the TSG augmentation is employed which alters the target reference to enforce constraints. Simulation results demonstrate the effectiveness of the TSG-guided MPC-CBF approach.

\end{abstract}

\section{INTRODUCTION}
\input{_I.Introduction/intro}

\section{PRELIMINARIES \label{sec:preliminaries}}
\subsection{System Description}
\input{_II.Preliminaries/a_dynamics}

\subsection{Control Barrier Function \label{subsec:cbf}}
\input{_II.Preliminaries/c_cbf}
\subsection{Collision Cone CBF (C3BF) \label{subsec:c3bf}}
\input{_II.Preliminaries/h_c3bf}

\section{PROBLEM FORMULATION \label{sec:problem}}
\input{_III.Problem/_intro}

\section{METHODOLOGY \label{sec:method}}
\subsection{System Dynamics \label{subsec:system-dynamics}}
\input{_IV.Methodology/a_dynamics}
\subsection{Time Shift Governor\label{subsec:tsg}}
\input{_IV.Methodology/g_tsg}
\subsection{CBF for Adaptive Cruise Control (ACC) \label{subsec:cbf_constraints}}
\input{_IV.Methodology/c_cbf}
\subsection{Relaxed C3BF for Dynamic Obstacle Avoidance \label{subsec:c3bf-constraints}}
\input{_IV.Methodology/h_c3bf}
\subsection{TSG-Guided MPC-CBF Optimization Problem \label{subsec:optimization-problem}}
\input{_IV.Methodology/f_mpccbf}

\section{RESULTS \label{sec:experiments}}
\input{_V.Experiments/_intro}
\subsection{Simulation Specifications}
\input{_V.Experiments/a_sim_spec}
\subsection{Simulation Results}
\input{_V.Experiments/b_result}

\section{CONCLUSION \label{sec:conclusion}}
\input{_VI.Conclusion/conclusion}


\addtolength{\textheight}{-12cm}   






\bibliographystyle{IEEEtran}
\typeout{}
\bibliography{references.bib}

\end{document}

%% file: _I.Introduction/intro.tex
Among the various subsystems that enable autonomous driving, Adaptive Cruise Control (ACC) plays a significant role in managing vehicle speed and maintaining safe distances from other vehicles. 
Various approaches for ACC have been developed to reduce the driver's workload, enhance driving safety, improve energy efficiency, and traffic flow \cite{yu2022researches}. 
Compared to manually driven vehicles, studies have shown that well-designed ACC systems reduce collision risks while improving road safety \cite{li2017evaluating}. However, ACC performance can be degraded in dynamically changing environments, where constraints are time-dependent and rapidly-varying. These environments involve unpredictable lead vehicle maneuvers and dynamically moving obstacles, beyond the structured flow of conventional traffic.

Model Predictive Control (MPC) uses the solution to a constrained optimal control problem (OCP) defined over a finite horizon to enforce constraints~\cite{del2010automotive}. Notably, reference~\cite{corona2008adaptive} presents that MPC-based ACC methods can outperform non-MPC approaches in dealing with nonlinearities. The advanced MPC designs integrate mode selection layers and adaptive control strategies to adjust MPC parameters in real-time \cite{naus2010model,zhang2018integrated}. 

Recently, the integration of Control Barrier Functions (CBFs) with MPC has gained attention. MPC-CBF approaches \cite{zeng_safetycritical_2021} utilize the solution to OCP with constraints that dictate that the system is within a forward invariant safe set. The CBFs provide a mathematical framework to keep the system state within a defined safe set \cite{ames2019control}. 
However, finding a valid CBF is not straightforward. In particular, selecting appropriate parameters for the class \( \mathcal{K} \) function in the CBF inequality is crucial for ensuring the feasibility of the safety constraints \cite{garg2024advances}. A recent work~\cite{kim2024learning} proposed a learning method to adapt class \( \mathcal{K} \) functions, however, it requires extensive offline trajectory sampling and training to ensure reliability. 
Moreover, CBFs require additional enhancements to enforce the safety constraints when safety constraints are time-varying. In such cases, maintaining recursive feasibility becomes challenging. For instance, the Collision Cone CBF (C3BF)~\cite{tayal2024collision} incorporates relative velocity to avoid dynamic obstacles but this approach can be too conservative and/or often fails to maintain feasibility. Motivated by these considerations, in this paper we propose a relaxed version of C3BF with slack variables, allowing soft constraint violations under dynamic uncertainty. By penalizing these slack variables in the cost function, the controller minimizes violations, thereby maintaining practical safety guarantees.

However, even with relaxed constraints, the challenge of ensuring safety when reference trajectories are varying, e.g., due to sudden changes in lead vehicle motion, remains unsolved. This motivates our augmentation of the Time Shift Governor (TSG) and a TSG-guided MPC-CBF architecture. The TSG is an add-on scheme that enforces constraints in a nominal closed-loop system. As other parameter governors~\cite{kolmanovsky2006parameter}, the TSG has low computational complexity since the optimization of only a scalar parameter is necessary to enforce safety constraints. The use of TSG has been considered for spacecraft formation flying~\cite{frey2016time}, spacecraft rendezvous in the Earth-Moon system~\cite{kim2024timeACC}, and along Earth orbits~\cite{kim2024timeAIAA, kim2025constrained}. But its potential for robust real-time safety enforcement in ground vehicle scenarios remains unexplored.

In this paper, we consider the integration of TSG into MPC-CBF formulations for ACC in dynamic environments with rapidly moving obstacles (see Fig.~\ref{fig:overview} for an overview).
Although our approach introduces an extra decision variable into the optimization problem, we demonstrate that this improves robustness in ensuring constraint satisfaction and maintaining recursive feasibility when the lead vehicle and obstacles motion is rapidly changing.
Furthermore, we extend the standard C3BF by incorporating slack variables, which balance the feasibility of safety constraint enforcement under dynamic conditions. 

The original contributions of this paper are:
\begin{itemize}
    \item We propose an MPC-CBF formulation for ACC, which leverages discrete-time CBFs~\cite{agrawal_discrete_2017} to enforce safety constraints.    
    \item Inspired by~\cite{alvarado2007robust}, we propose a novel framework that integrates the TSG into the MPC-CBF formulation, enhancing constraint enforcement capabilities in dynamic and uncertain environments.    
    \item We enhance the standard C3BF by introducing slack variables, improving feasibility and responsiveness in environments with moving obstacles.
    \item The effectiveness of the proposed approach is validated through simulation studies, which incorporate dynamic obstacles and time-varying lead vehicle velocities in low speed driving (e.g., city streets and parking lots).
\end{itemize}

This paper is organized as follows: Section~\ref{sec:preliminaries} presents the preliminaries, including the system description and definition of CBF. In Section~\ref{sec:problem}, we formulate the problem statement. In Section~\ref{sec:method}, we describe our proposed methodology, including the MPC-CBF framework and TSG implementation. In Section~\ref{sec:experiments}, we present the simulation results and discussion. Finally, Section~\ref{sec:conclusion} concludes the paper.

%% file: _II.Preliminaries/a_dynamics.tex
We consider a class of systems represented by a discrete-time nonlinear model of the form,
\begin{equation}
\label{eq:system_dynamics}
    x_{k+1} = f(x_k, u_k),
\end{equation}
where \( x_k \in \mathcal{X} \subset \mathbb{R}^n \) is the state vector at time step $k$, \( u_k \in \mathcal{U}  \subset \mathbb{R}^m \) is the control input at time step k, and \( f : \mathbb{R}^n \times \mathbb{R}^m \to \mathbb{R}^n \) is a continuously differentiable function representing the system dynamics which is locally Lipschitz. Here, $\mathcal{X}$ and $\mathcal{U}$ represent the admissible state and control input sets, respectively. In this work, we assume that the state $x$ at time step $k$ can be fully measured.

%% file: _II.Preliminaries/c_cbf.tex
Control Barrier Functions (CBFs) enforce safety constraints by maintaining the system state within a safe set. 
For a given safety constraint \( h(x) \geq 0 \), the discrete-time CBF is defined as
\begin{equation}
    \psi_0(x) = h(x),
\end{equation}
where \( h(x) \) is a continuously differentiable function. The CBF condition ensures the system remains within the safe set, i.e., \( \mathcal{C} = \{ x \in \mathbb{R}^n \mid h(x) \geq 0 \} \).

\begin{definition}[Forward Invariant Set]
A set \( \calC \subset \mathbb{R}^n \) is said to be forward invariant for the system described by \eqref{eq:system_dynamics} if, for any initial state \( \vx_0 \in \calC \) and a specified control input \(u_k = u(x_k) \), the state trajectory \( \vx_t \) remains within \( \calC \) for all future time steps, \( k \geq 0 \).
\end{definition}

\begin{definition}[Discrete-Time Control Barrier Function] \label{def:dt_cbf}
Consider a discrete-time system \eqref{eq:system_dynamics}. A continuous function \( h: \mathbb{R}^n \rightarrow \mathbb{R} \) is a Discrete-Time CBF for the system if there exists a class \( \calK \) function \( \alpha \), where $\alpha$ should satisfy $\alpha(\gamma) < \gamma$ for all $\gamma > 0$, and a control function, \(u_k = u(x_k) \), such that:
\begin{equation} \label{eq:dt_cbf_condition}
    h(x_{k+1}) - h(x_k) \geq -\alpha(h(x_k)),\; \forall x_k \in \mathbb{R}^n,\; \forall k \geq 0.
\end{equation} 

This condition ensures that the state \( \vx_k \) can be maintained within the safe set 
\( \mathcal{C} \) for all future time steps.
The proof is similar to the proof of \cite[Theorem 1]{ahmadi2019safe}. 
\end{definition}

\begin{remark} \label{rem:class_k}
The class \( \mathcal{K} \) function \( \alpha \) in \eqref{eq:dt_cbf_condition} can be chosen as \( \alpha(r) = \gamma r \) with \( 0 < \gamma \leq 1 \). This linear choice of \( \alpha \) ensures that \( h(x_k) \geq (1 - \gamma)^k h(x_0) \) for all \( k \in \mathbb{N} \), thereby guaranteeing the forward invariance of \( \mathcal{C} \) \cite{agrawal_discrete_2017}.
\end{remark}

%% file: _II.Preliminaries/h_c3bf.tex
The Collision Cone Control Barrier Functions (C3BFs) in~\cite{tayal2024collision} ensure that the relative velocity vector between the ego vehicle and a moving obstacle does not point into a geometrically defined collision cone. 
The C3BFs guarantee safety by constraining the direction of approach.
The obstacle is approximated as an ellipse, and a collision cone is formulated using tangents from the ego vehicle’s center. 
The collision risk increases when the relative velocity lies within the collision cone. The C3BF has the following form,
\begin{equation}
h_{\text{C3BF}}(x_k) = \langle p_{\text{rel}}, v_{\text{rel}} \rangle + \|p_{\text{rel}}\| \cdot \|v_{\text{rel}}\| \cdot \cos \phi,
\end{equation} \label{eq:c3bf}
where \( p_{\text{rel}} \) is the relative position vector between the vehicle and the obstacle, \( v_{\text{rel}} \) is the relative velocity, \( \cos \phi = \sqrt{1 - \frac{r^2}{\|p_{\text{rel}}\|^2}} \), and \( r \) is the safety radius.

The constraint \(h_{C3BP} \geq 0\) ensures that the angle between \( p_{\text{rel}} \) and \( v_{\text{rel}} \) is greater than \( \phi \), avoiding collisions. The resulting set \( \mathcal{C}_{\text{C3BF}} = \{x_k : h_{\text{C3BF}}(x_k) \geq 0 \} \) is forward invariant when enforced using a discrete-time CBF constraint of the form:
\begin{equation}
h_{\text{C3BF}}(x_{k+1}) - h_{\text{C3BF}}(x_k) + \alpha(h_{\text{C3BF}}(x_k)) \geq 0.
\end{equation} \label{eq:c3bf_constraint}
The theoretical validity of this constraint for kinematic bicycle models is proven in~\cite[Theorem 2]{tayal2024collision}.

%% file: _III.Problem/_intro.tex
We address the problem of safely steering a vehicle within an environment $\mathcal{W}$ containing known obstacles and a lead vehicle. The vehicle's dynamics are given by the discrete-time model \eqref{eq:system_dynamics}.


Our goal is to design an ACC controller that enables the ego vehicle to (i) track a curved road trajectory, (ii) maintain a safe distance from a potentially non-cooperative lead vehicle, (iii) avoid collisions with dynamic obstacles, and (iv) adapt in real time to environmental variations. A key requirement is to guarantee the recursive feasibility of safety constraints.

Given the current state \(x_k\), the goal is to compute a sequence of control inputs \(u_{0:N-1}\) and a time shift parameter \(\tau_{\tt shift}\) over a finite prediction horizon \(N\), such that:
 
\begin{enumerate}
    \item The ego vehicle follows a nominal reference trajectory defined by the road centerline.
    \item The safety constraints—such as minimum distance to the lead vehicle and CBF-based obstacle avoidance—are satisfied at each step.
    \item A relaxed version of the collision cone CBF is used to improve feasibility in dynamic obstacles.
    \item The TSG dynamically adjusts the lead-vehicle reference using \(\tau_{\tt shift}\), enhancing recursive feasibility without extending the prediction horizon.
\end{enumerate}

The constraints and cost terms are formalized in the optimization problem described in Section~\ref{subsec:optimization-problem}. Our formulation extends standard MPC-CBF by introducing \(\tau_{\tt shift}\) as an additional decision variable and relaxing the CBF conditions via slack variables, thus improving robustness to dynamic uncertainties without increasing computational complexity significantly.




%% file: _IV.Methodology/a_dynamics.tex
We adopt a kinematic bicycle model~\cite{tayal2024collision} with vehicle parameters from the 2016 KIA Soul~\cite{oh2021handling, kiasoul2016specs}.

The vehicle state is defined as \( x_k = [x_k, y_k, \theta_k, v_k]^\top \), where \( (x_k, y_k) \) denote the position, \( \theta_k \) is the heading angle, and \( v_k \) is the longitudinal velocity at time step \( k \). The control input is \( u_k = [a_k, \beta_k]^\top \), where \( a_k \) is the longitudinal acceleration and \( \beta_k \) is the vehicle's slip angle at the center of mass. Here, the slip angle is $\beta = \tan^{-1} \left( \frac{l_r}{l_f + l_r} \tan(\delta) \right)$, where $\delta$ is the steering angle of the vehicle, while $l_f$ and $l_r$ are the distances of the front and rear axle from the center of mass, respectively.

Under the small-angle assumption, \( \cos\beta \approx 1, \sin\beta \approx \beta \), the continuous-time dynamics can be rewritten in control-affine form:
\begin{equation}
\dot{x} = f(x) + g(x) u =
\begin{bmatrix}
v \cos \theta \\
v \sin \theta \\
0 \\
0
\end{bmatrix}
+
\begin{bmatrix}
0 & -v \sin \theta \\
0 & v \cos \theta \\
0 & \frac{v}{l_r} \\
1 & 0
\end{bmatrix}
\begin{bmatrix}
a \\
\beta
\end{bmatrix}.
\end{equation}  \label{eq:kinematic_bicycle}

They are converted to discrete-time using Euler integration.

%% file: _IV.Methodology/g_tsg.tex
\begin{figure}[tbp]
\centering
\includegraphics[width=0.99\linewidth]{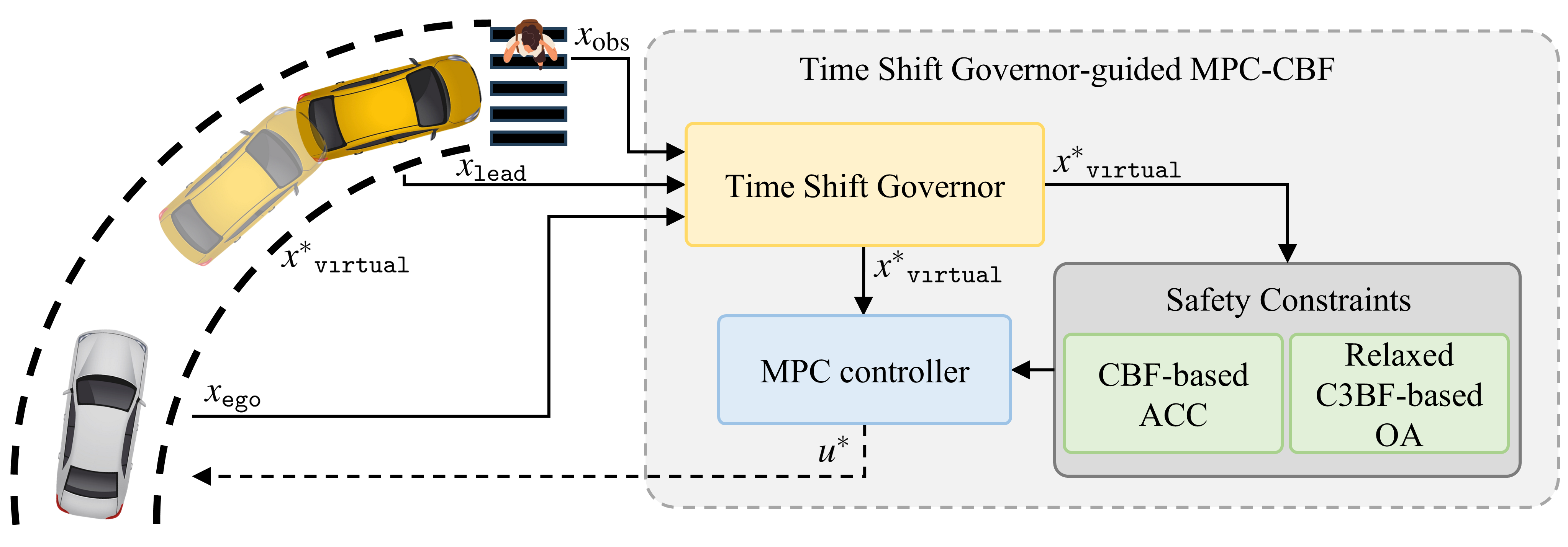}
\caption{Overview of the control system architecture of TSG-guided MPC-CBF. A detailed explanation is provided in the methodology section.}
\label{fig:overview}
\end{figure}

The Time Shift Governor is an add-on scheme to the nominal controller that enforces constraints by using a time-shifted reference, determined by a decision variable \(\tau_{\tt shift}\)~\cite{frey2016time}.
In this work, we augment the TSG to help enforce constraints when unexpected behaviors of the lead vehicle or changes in the environment occur.  

In our setting, the TSG dynamically generates a virtual reference target, corresponding to a shifted version of the lead vehicle’s trajectory. Specifically, \(\tau_{\tt shift}\) shifts the reference trajectory backward in time, enabling the ego vehicle to respond as if the lead vehicle were in an earlier position. 
This reference adaptation enhances the recursive feasibility of the control problem without the need to modify the original prediction model, thereby enhancing constraint satisfaction in dynamic environments.

The time shift is selected as the smallest value in magnitude that ensures the safety constraints are enforced. The time-shifted position of the lead vehicle is projected onto a predefined reference trajectory to generate the virtual command.

The virtual target is calculated as:
\[
x_{\tt{virtual}}(\tau) = f_{\tt{shift}}(\theta_{\tt{lead}}, \tau, \tau_{\tt{shift}}, r_{\tt{c}}){\color{blue},}
\]
where the function \(f_{\tt{shift}}(\theta_{\tt{lead}}, \tau, \tau_{\tt{shift}}, r_{\tt{c}})\) is defined as
\begin{align}
    f_{\tt{shift}}(\theta_{\tt{lead}}, \tau, \tau_{\tt{shift}}, r_{\tt{c}}) =
    r_{\tt{c}}
    \begin{bmatrix} 
        \cos \theta_{\tt{lead}}(\tau) \\[4pt] 
        \sin \theta_{\tt{lead}}(\tau) 
    \end{bmatrix}{\color{blue}.}\notag
\end{align}

The virtual heading angle is then given by
\begin{equation}
    \theta_{\tt{lead}}(\tau) = \text{atan2} \left(y_{\tt{lead}}(\tau + \tau_{\tt{shift}}), x_{\tt{lead}}(\tau + \tau_{\tt{shift}}) \right), \notag
\end{equation}
where \( (x_{\tt{lead}}, y_{\tt{lead}}) \) represents the $x$-$y$ coordinates of the lead vehicle at time instant \( \tau \) and \( r_{\tt{c}} \) denotes the radius of the circular reference trajectory onto which the virtual vehicle is projected.

By using the TSG, we introduce an additional decision variable \(\tau_{\tt{shift}}\) into the optimization problem in the next section. However, this additional decision variable enhances robustness to lead vehicle deviations and helps maintain recursive feasibility (especially, when short horizon MPC-CBF is used). In particular, it allows constraint satisfaction to be preserved under aggressive or non-predictable maneuvers, which would otherwise render the MPC-CBF problem infeasible.

%% file: _IV.Methodology/c_cbf.tex
To ensure safety with respect to the lead vehicle, we define a CBF that maintains a safe distance between the ego and lead vehicles.
For simplicity, we adopt the guideline from~\cite{vogel2003comparison}, which suggests that the minimum safe distance should be half the speedometer reading, i.e., $D \geq \frac{v}{2}$, where \( D \) is the distance in meters and \( v \) is the velocity in kilometers per hour. As in~\cite{ames2019control}, we employ the following CBF to ensure a safe distance to the leading vehicle:
\begin{equation} \label{eq:cbf_constraint_lead}
h_{\tt{virtual}}(x_{\tt{ego}}, x_{\tt{virtual}}) = \|p(x_{\tt{ego}}) - p(x_{\tt{virtual}})\| - r_{\ell} - 1.8v,
\end{equation}
where \( r_{\ell} \) is the combined radius for the ego and the leading vehicle, the factor 1.8 converts the units to meters and seconds, and \( v \) is the ego vehicle's velocity. The term $p(x_{\tt{virtual}})$ is used as a reference trajectory for the lead vehicle; note that it depends on the virtual target.

However, since \eqref{eq:cbf_constraint_lead} only constrains the ego vehicle relative to the virtual target, it does not directly prevent collisions with the lead vehicle, particularly when the virtual target is positioned ahead of the lead vehicle. To address this, we introduce an additional constraint ensuring that the ego and lead vehicles remain physically separated:
\begin{equation} \label{eq:cbf_distance_constraint}
    \|p(x_{\tt{ego}}) - p(x_{\tt{lead}})\| \geq r_{\tt{ego}} + r_{\tt{lead}}.
\end{equation}
This ensures that the ego vehicle maintains a safe physical distance from the lead vehicle.

%% file: _IV.Methodology/h_c3bf.tex
To handle dynamic obstacles while avoiding overly conservative reactions, we adopt a relaxed version of the C3BF~\cite{tayal2024collision}. The standard C3BF defines a forward-invariant safe set by ensuring that the relative velocity vector remains outside a collision cone defined by the relative position.

However, in highly dynamic environments, strict enforcement can lead to infeasibility, especially when the obstacle is far but moving toward the ego vehicle. To mitigate this, we introduce a non-negative slack variable \( \delta_{\text{C3BF}} \geq 0 \) into the discrete-time CBF condition:
\begin{equation}
h_{\text{C3BF}}(x_{k+1}) - h_{\text{C3BF}}(x_k) + \alpha(h_{\text{C3BF}}(x_k)) - \delta_{\text{C3BF}} \geq 0.
\end{equation}
The slack variable acts as an auxiliary input to the MPC problem, penalized in the cost function via a linear term \( \lambda \delta_{\text{C3BF}} \), where \( \lambda > 0 \) controls the softness of the constraint. Box constraints are enforced to limit its impact: $0 \leq \delta_{\text{C3BF}} \leq \bar{\delta}.$
The underlying C3BF expression \( h_{\text{C3BF}} \) remains identical to the formulation introduced in Section~\ref{subsec:c3bf}.

%% file: _IV.Methodology/f_mpccbf.tex
To adapt to dynamically evolving environments, we incorporate a time shift parameter \( \tau_{\text{shift}}\) into the MPC-CBF approach from~\cite{zeng_safetycritical_2021}, which alters the reference trajectory of the lead vehicle. We also utilize a relaxed C3BF constraint with a slack variable to improve recursive feasibility when there are moving obstacles.

The resulting MPC problem is formulated as: 
\begin{subequations} \label{eq:mpc-cbf-formulation}
\begin{align}
\min_{\substack{
    u_{0:N-1}, \\
    \delta_{\text{C3BF},0:N-1}
}}
\quad & \| r_{\tt{ego}, N} - r_{\text{ref}} \|^2_{Q} + \sum_{k=0}^{N-1} \Big( \| r_{\tt{ego}, k} - r_{\text{ref}} \|^2_{Q} \notag \\
& \quad + u_k^T R u_k + \lambda \delta_{\text{C3BF}, k} + \| v_k - v_{\text{des}} \|^2_{Q_d} \notag \\
& \quad + \| \tau_{\tt{shift}} \|^2_{Q_{\tt{shift}}} \Big) \label{eq:mpc-cbf-finalcost} \\
\text{s.t.} \quad & x_{k+1} = f(x_k, u_k), \\
& x_k \in \mathcal{X}, \quad u_k \in \mathcal{U}, \quad \delta_{\text{C3BF}, k} \in [0, \bar{\delta}], \\
& x_{\tt{virtual}}(\tau) = f_{\tt{shift}}(\theta_{\tt{lead}}, \tau_{k}, \tau_{\tt{shift}}, r_{\tt{c}}), \label{eq:mpc-cbf-tsg}\\
& h_{\tt{virtual}, {k+1}} + (\gamma_{\tt{virtual}} - 1) h_{\tt{virtual}, {k}} \geq 0, \label{eq:mpc-cbf-cbfconstraintlead} \\ 
& \|p(x_{\tt{ego}}) - p(x_{\tt{lead}})\| \geq r_{\tt{ego}} + r_{\tt{lead}}, \label{eq:mpc-cbf-ego-lead-distance} \\
& h_{\text{C3BF}}(x_{k+1}) - h_{\text{C3BF}}(x_k) + \notag \\
& \quad  \alpha(h_{\text{C3BF}}(x_k)) \leq \delta_{\text{C3BF}, k} \quad \forall k = 0, \dots, N-1. \label{eq:mpc-cbf-relaxedc3bf}
\end{align}
\end{subequations}

Here, 
\(r_{\tt{ego},k}\) is the distance to the circular reference trajectory, \(r_{\text{ref}}\) is the radius, \(\tau_{\tt{shift}}\) is the time shift parameter introduced as a decision variable, \(u_k = [a_k, \beta_k]^T\) is the control input, \(v_{\text{des}}\) is the desired cruising velocity, and \(Q = Q^\top \succeq 0\), \(R = R^\top \succ 0\), \(Q_d = Q_d^{\sf T} \succ 0\), and \(Q_{\tt{shift}} = Q_{\tt{shift}}^{\sf T} \succ 0\) are weight matrices for the cost function. The variable \( \delta_{\text{C3BF}, k} \) is a slack introduced to relax the C3BF constraint, with upper bound \( \bar{\delta} \) and penalty coefficient \( \lambda \).

The virtual target $x_{\tt virtual}$ in \eqref{eq:mpc-cbf-tsg} is defined in Section~\ref{subsec:tsg} as a time-shifted projection of the lead vehicle’s position using $\tau_{\tt shift}$. This target is used in the virtual safety constraint $h_{\tt virtual,k}$ in \eqref{eq:mpc-cbf-cbfconstraintlead}, which is explicitly defined in \eqref{eq:cbf_constraint_lead}. The virtual target $x_{\tt virtual, k}$ is used to compute the projected position $p(x_{\tt{virtual},k})$, which directly depends on $\tau_{\tt shift}$ through the shifted heading angle and reference projection. The inclusion of $\tau_{\tt shift}$ as a decision variable enables the ego vehicle to anticipate delayed behavior from the lead vehicle and adapt accordingly, thereby enhancing recursive feasibility.

This formulation balances trajectory tracking, velocity maintenance, and control effort minimization while optimizing the time shift parameter considering (\ref{eq:mpc-cbf-tsg}). 
The overall formulation follows the general NMPC structure but incorporates both the TSG and CBF constraints, \eqref{eq:mpc-cbf-cbfconstraintlead} and \eqref{eq:mpc-cbf-relaxedc3bf}, to ensure safety. Additionally, the constraint \eqref{eq:mpc-cbf-ego-lead-distance} prevents potential collisions even when the virtual target is positioned ahead.

%% file: _V.Experiments/_intro.tex
Our proposed TSG-guided MPC-CBF framework has demonstrated its effectiveness in simulations on a circular path, which requires frequent turning maneuvers.
The results show that our approach enhances the robustness of ACC in dynamic environments, such as moving obstacles or sudden changes in the lead vehicle's behavior. By enabling real-time adaptation to environmental changes, our method significantly improves safety performance. 

\begin{figure*}[t]
\centering
\begin{subfigure}[t]{0.48\textwidth}
    \centering
    \begin{subfigure}[b]{0.49\textwidth} 
        \centering
        \includegraphics[width=\textwidth]{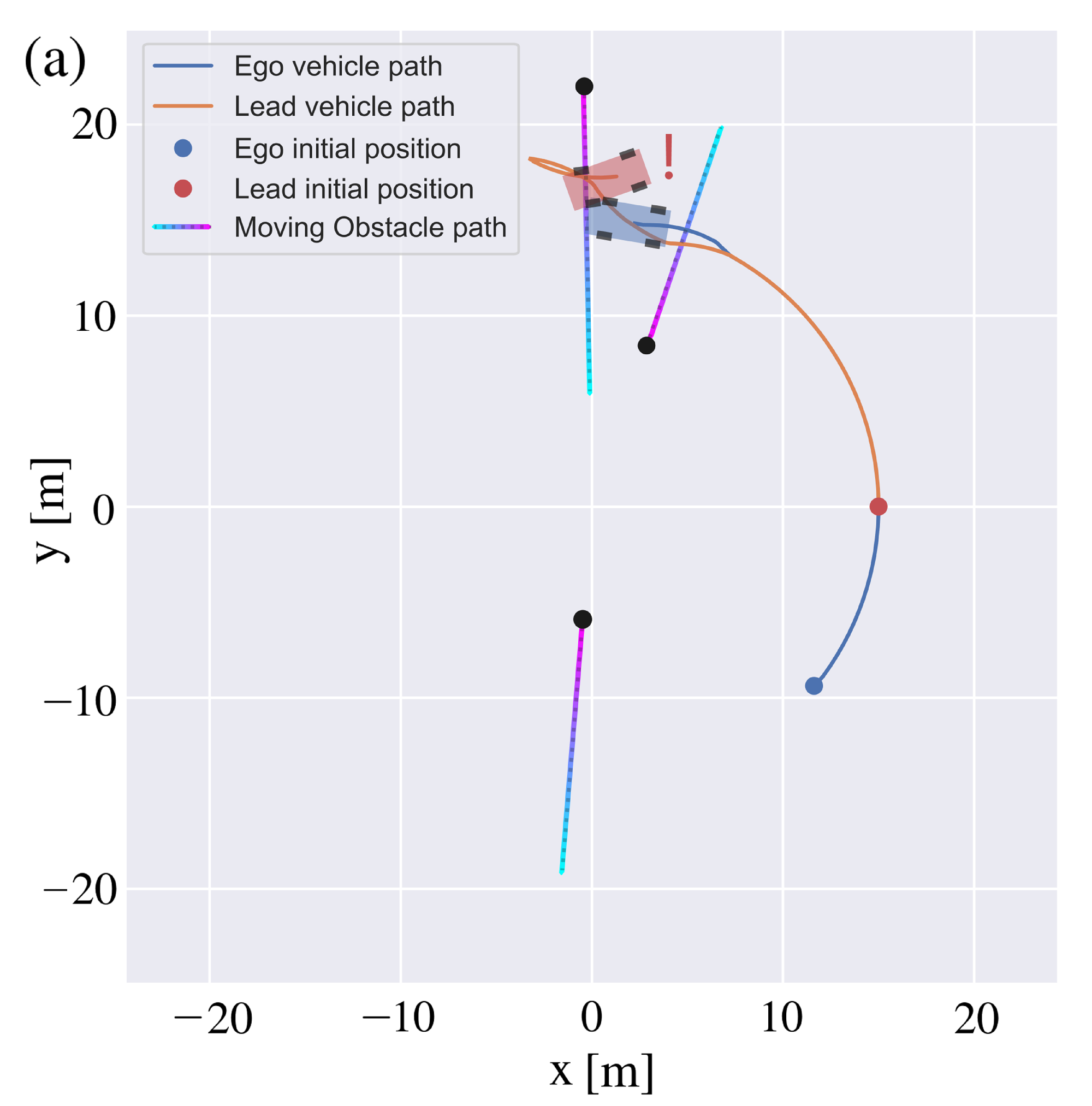}
        \label{fig:scen1_trajectories_1}
    \end{subfigure}\vspace{-1.5mm}
    \hfill
    \begin{subfigure}[b]{0.49\textwidth}
        \centering
        \includegraphics[width=\textwidth]{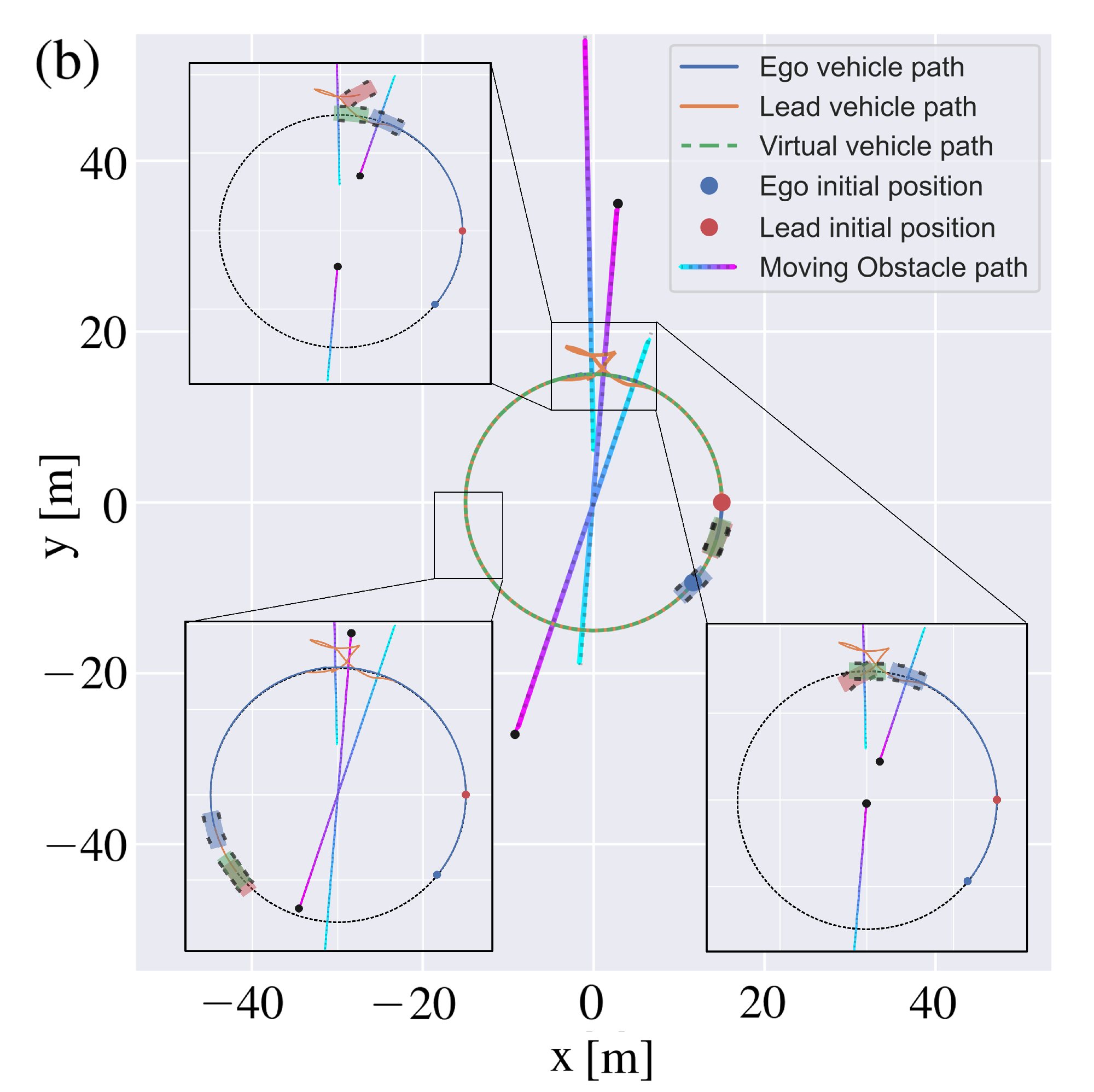}
        \label{fig:scen1_trajectories_2}
    \end{subfigure} \vspace{-5.0mm}
    
    \begin{subfigure}[b]{0.49\textwidth}
        \centering
        \includegraphics[width=\textwidth]{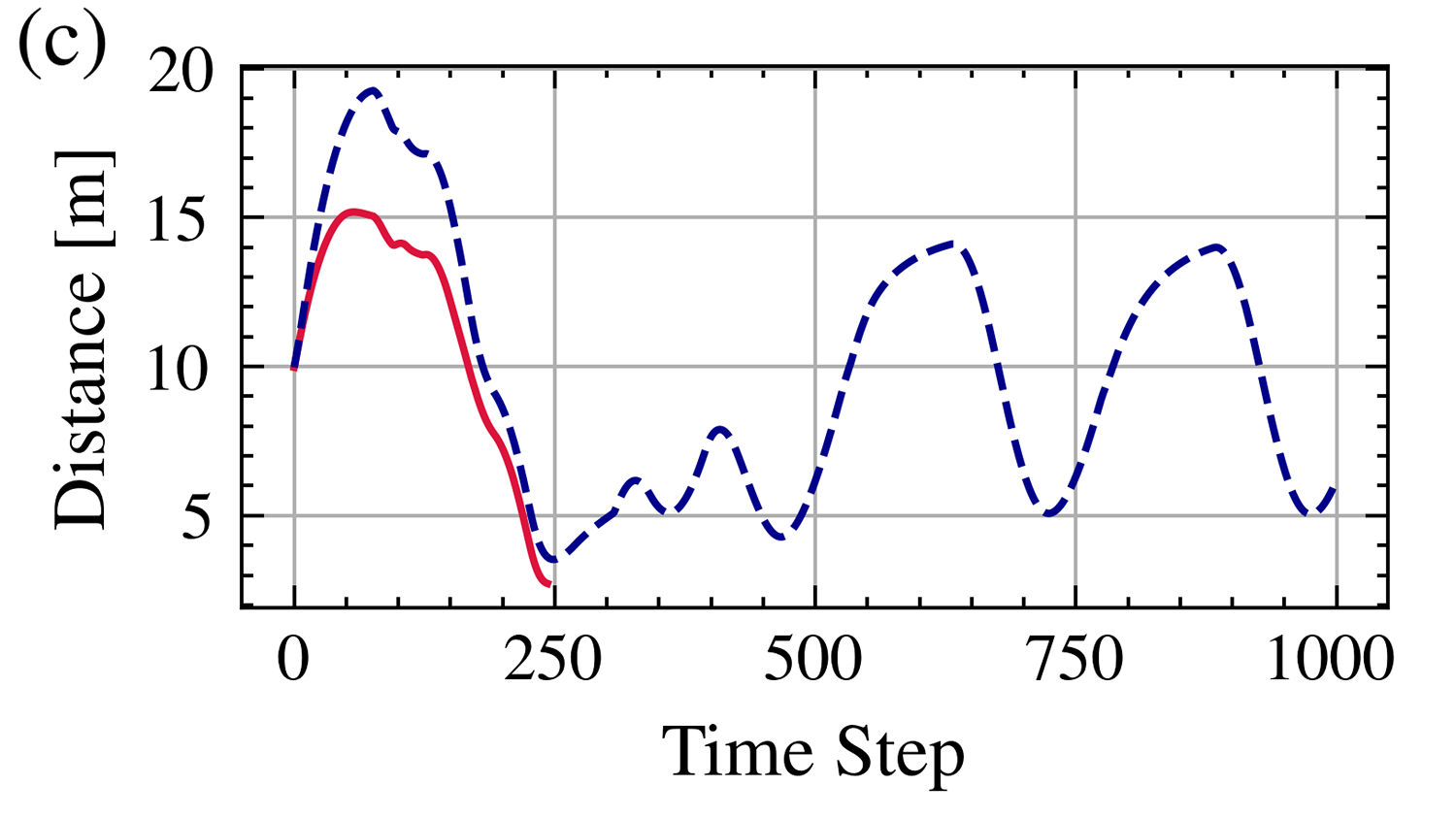}
        \label{fig:scn1_rel_dis}
    \end{subfigure}\vspace{-1.5mm}
    \hfill
    \begin{subfigure}[b]{0.49\textwidth}
        \centering
        \includegraphics[width=\textwidth]{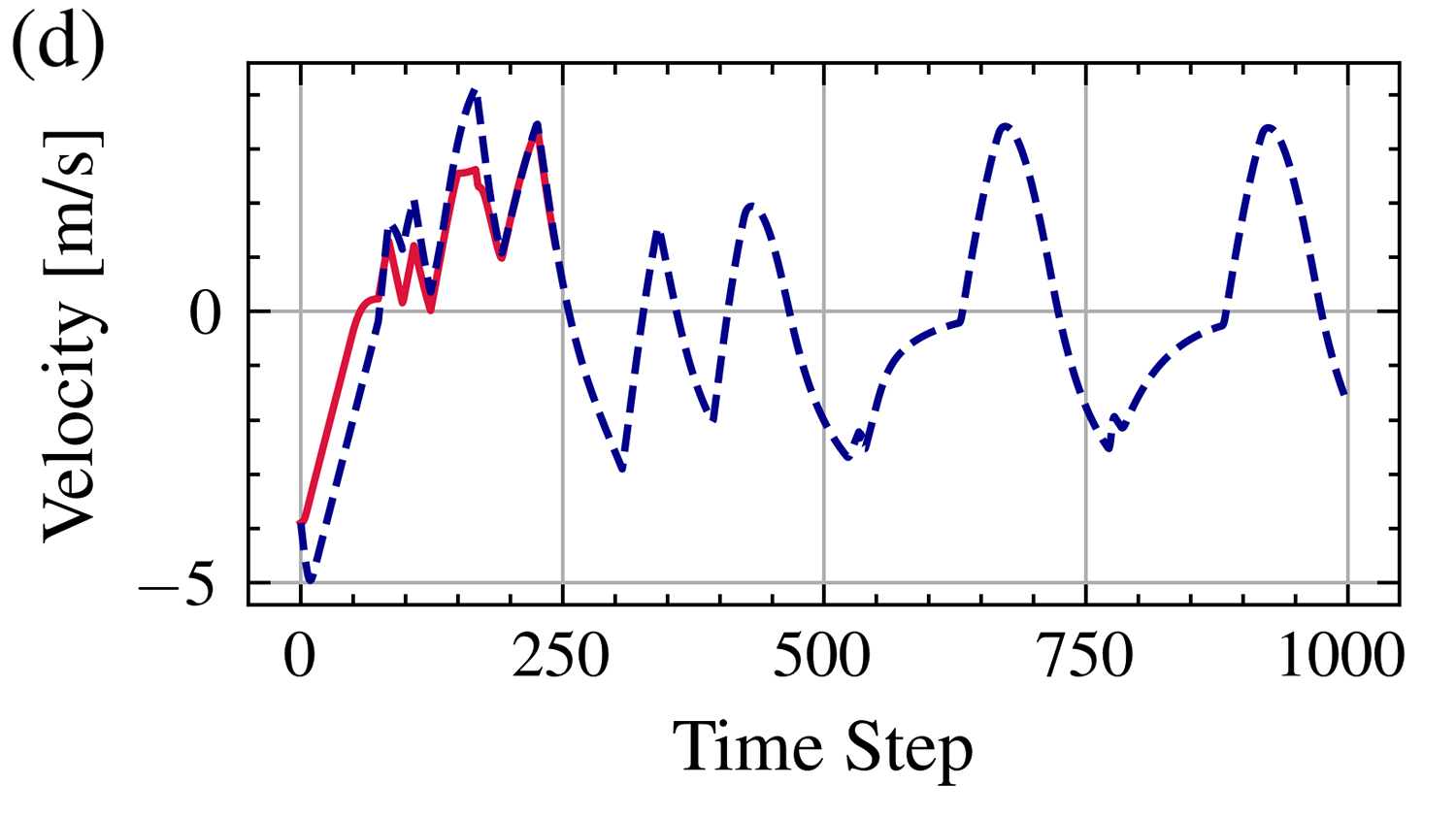}
        \label{fig:scn1_rel_vel}
    \end{subfigure} \vspace{-3.0mm}
    
    \begin{subfigure}[b]{0.49\textwidth}
        \centering
        \includegraphics[width=\textwidth]{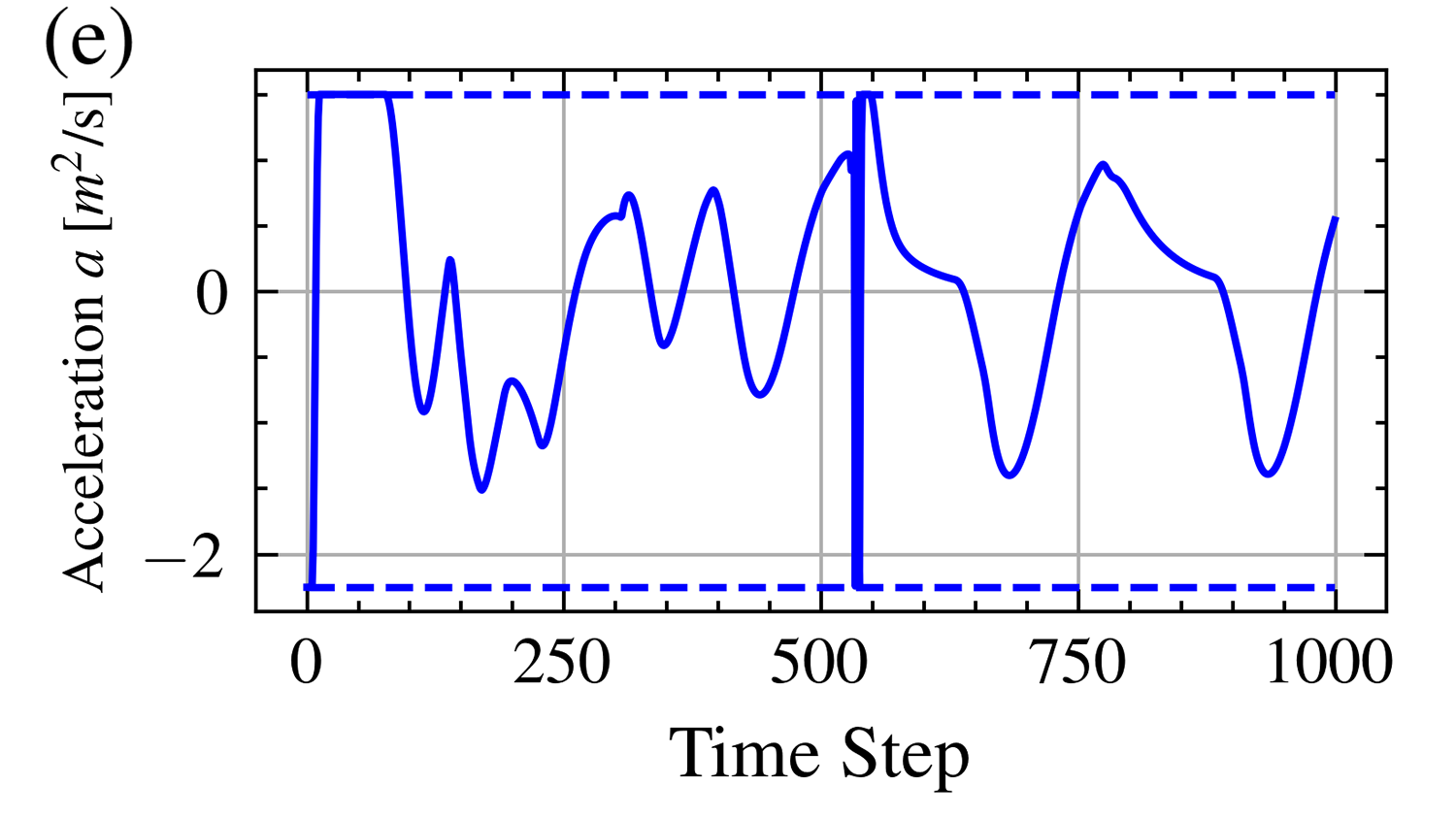}
        \label{fig:scn1_ctrl_acc}
    \end{subfigure}\vspace{-1.5mm}
    \hfill
    \begin{subfigure}[b]{0.49\textwidth}
        \centering
        \includegraphics[width=\textwidth]{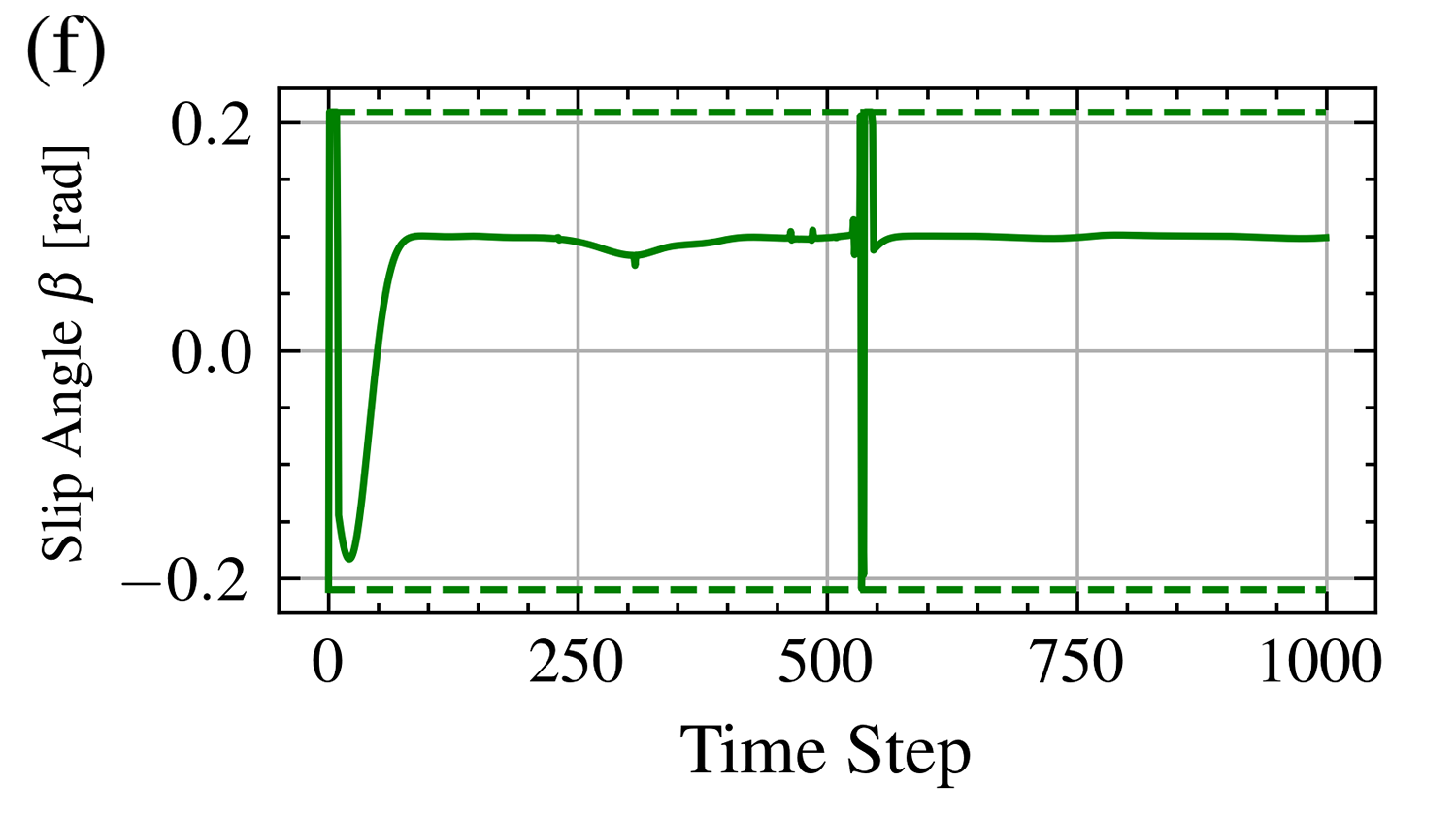}
        \label{fig:scn1_ctrl_omega}
    \end{subfigure} \vspace{-3.0mm}
    
    \begin{subfigure}[b]{0.50\textwidth}
        \centering
        \includegraphics[width=\textwidth]{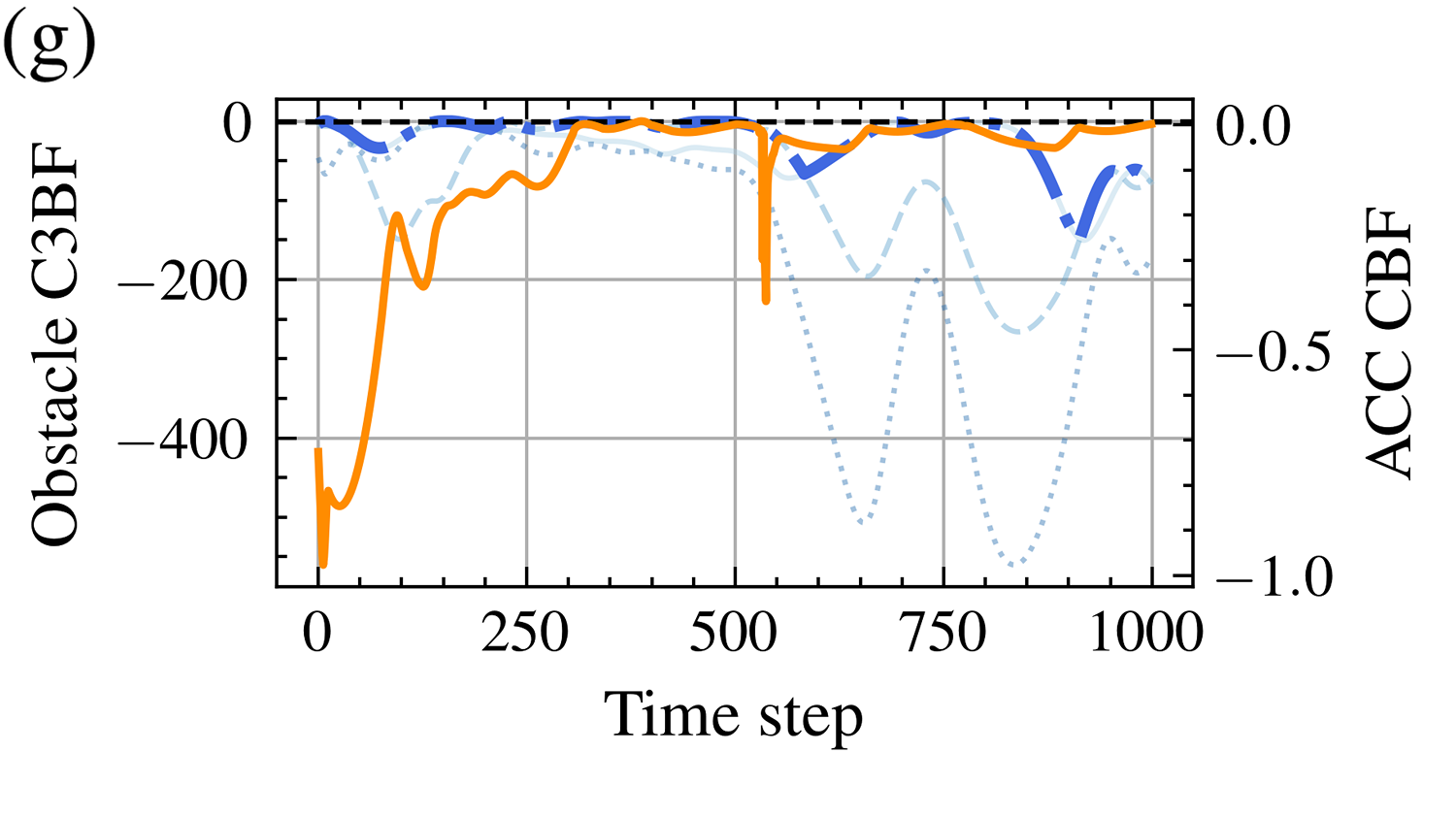}
        \label{fig:scn1_cbf}
    \end{subfigure}\vspace{-1.5mm}
    \hfill
    \begin{subfigure}[b]{0.48\textwidth}
        \centering
        \includegraphics[width=\textwidth]{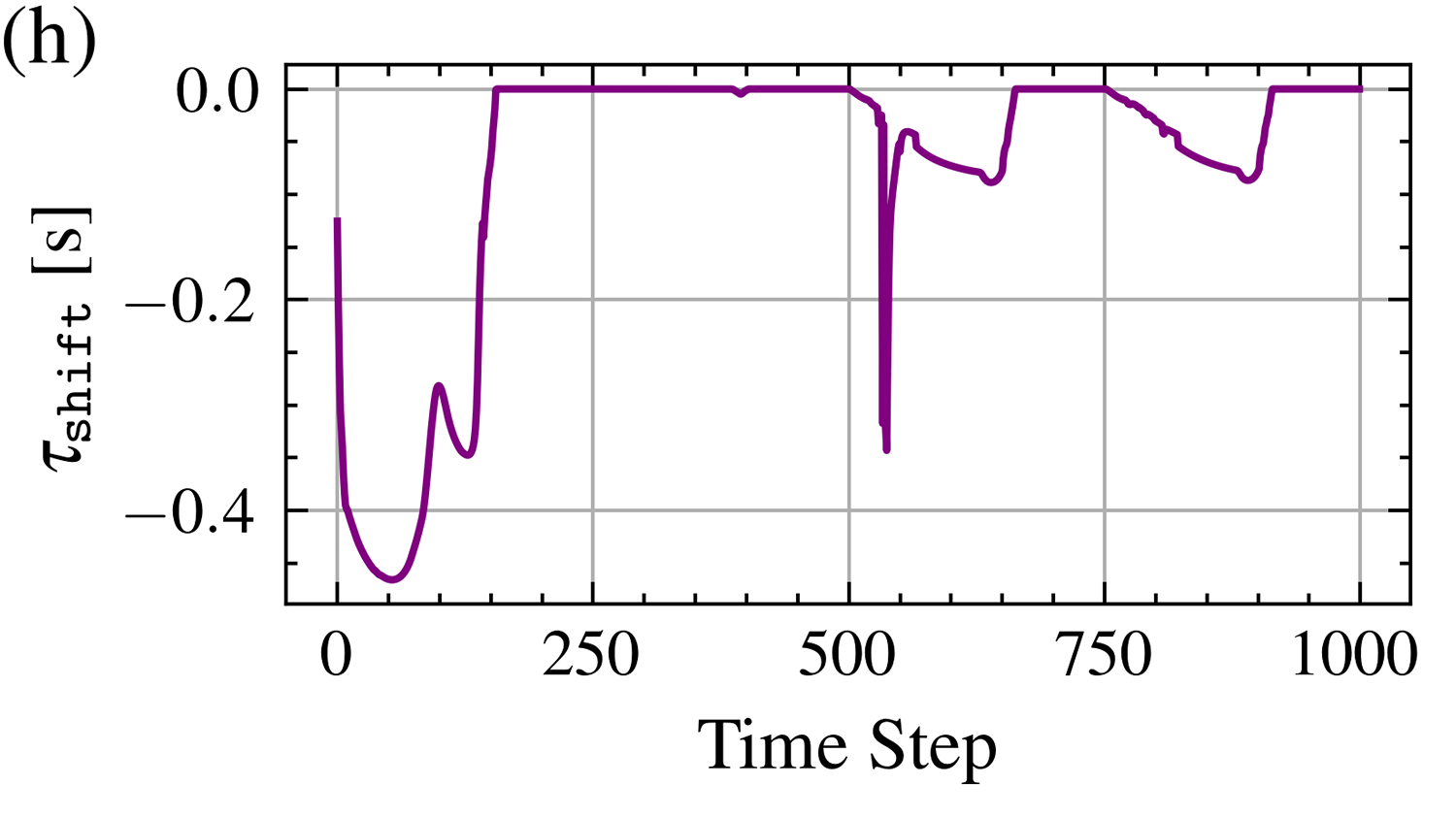}
        \label{fig:scn1_timeshift}
    \end{subfigure} \vspace{-6.0mm}
    \caption*{(A)} \label{fig:scen1}
\end{subfigure}
\hfill
\begin{subfigure}[t]{0.48\textwidth}
    \centering
    \begin{subfigure}[b]{0.49\textwidth}
        \centering
        \includegraphics[width=\textwidth]{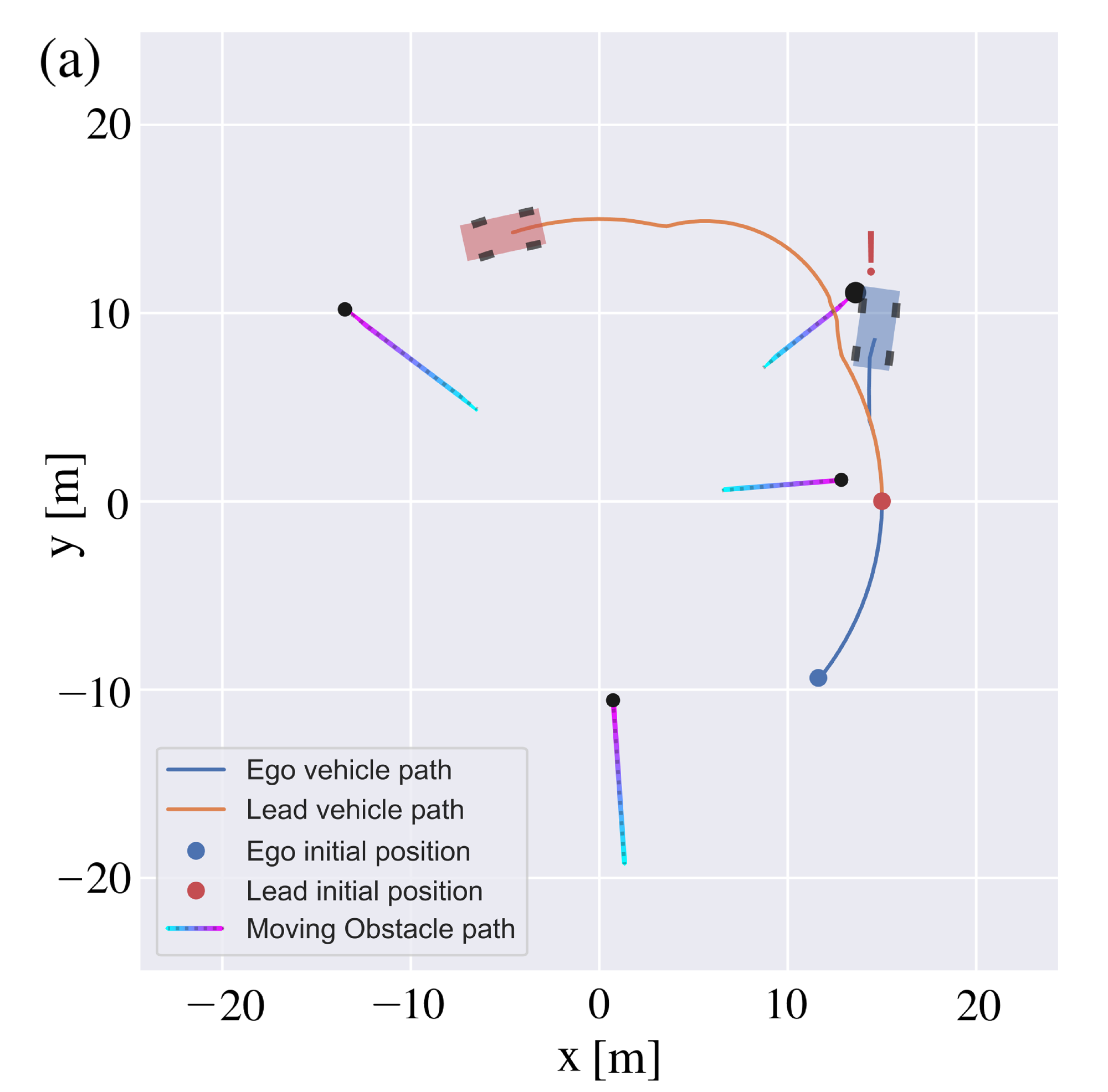}
        \label{fig:scen2_trajectories_1}
    \end{subfigure}\vspace{-1.5mm}
    \hfill
    \begin{subfigure}[b]{0.48\textwidth}
        \centering
        \includegraphics[width=\textwidth]{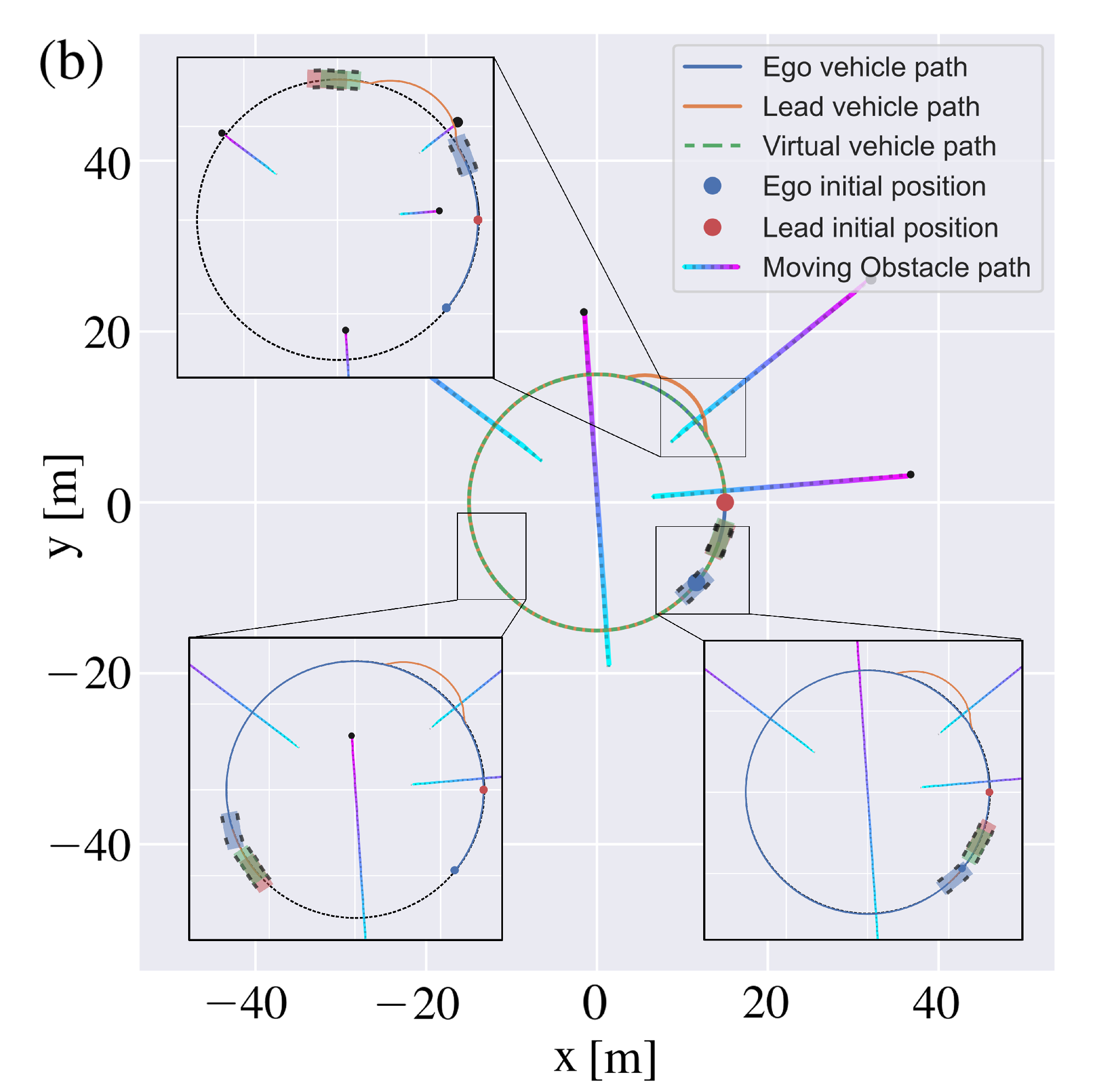}
        \label{fig:scen2_trajectories_2}
    \end{subfigure} \vspace{-0.5mm}
    
    \begin{subfigure}[b]{0.49\textwidth}
        \centering
        \includegraphics[width=\textwidth]{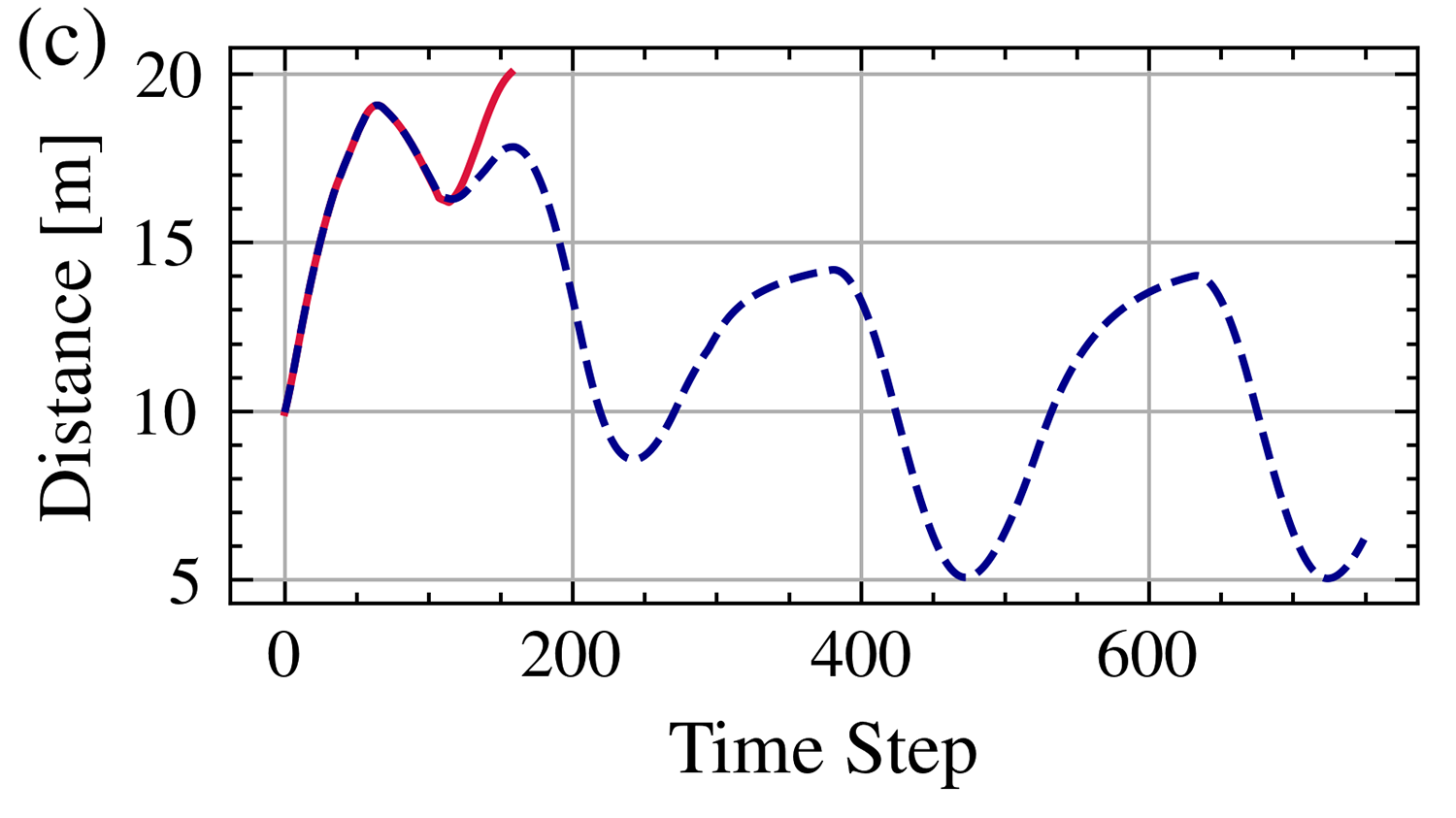}
        \label{fig:scn2_rel_dis}
    \end{subfigure}\vspace{-1.5mm}
    \hfill
    \begin{subfigure}[b]{0.49\textwidth}
        \centering
        \includegraphics[width=\textwidth]{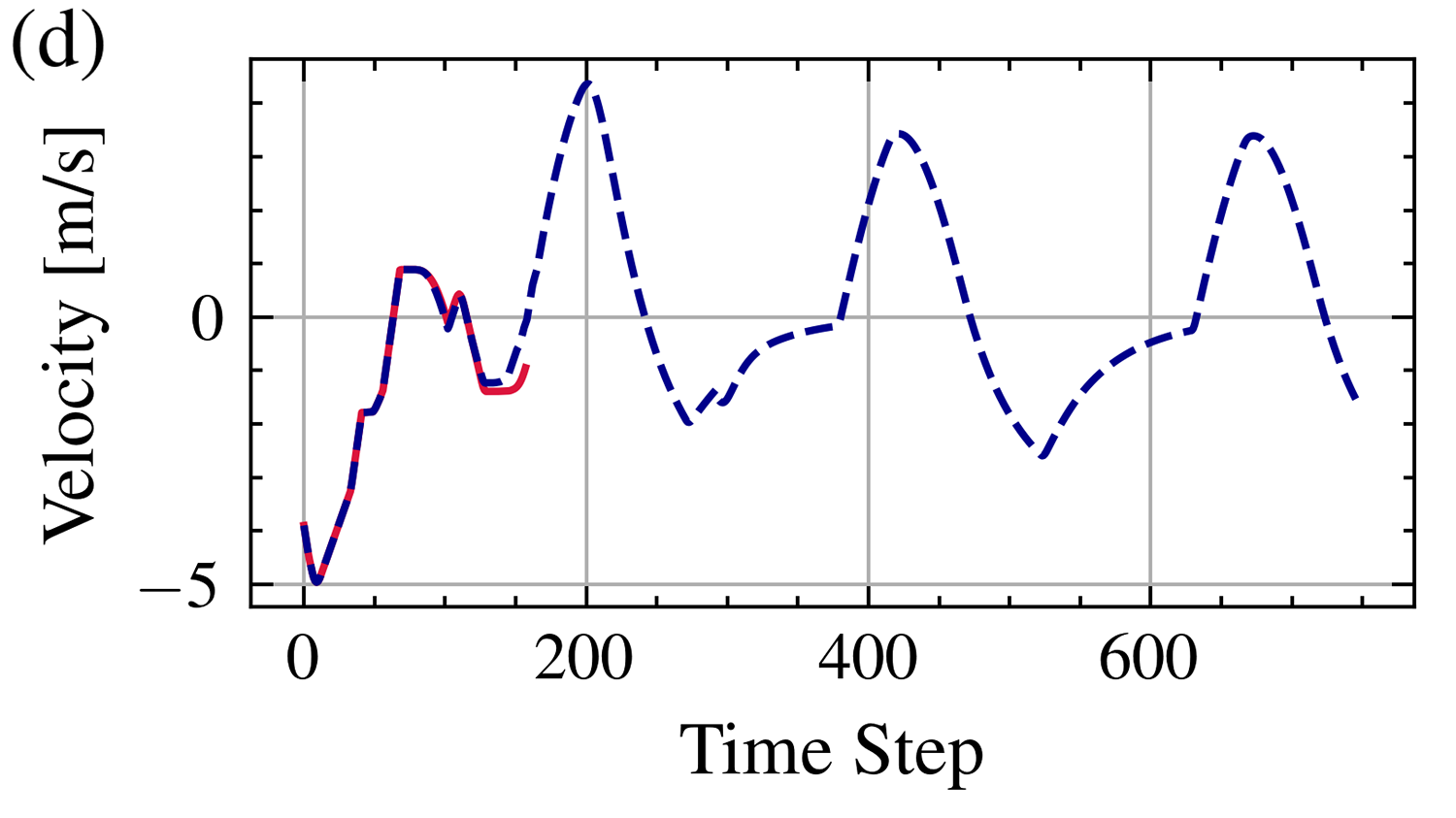}
        \label{fig:scn2_rel_vel}
    \end{subfigure} \vspace{-3.0mm}
    
    \begin{subfigure}[b]{0.49\textwidth}
        \centering
        \includegraphics[width=\textwidth]{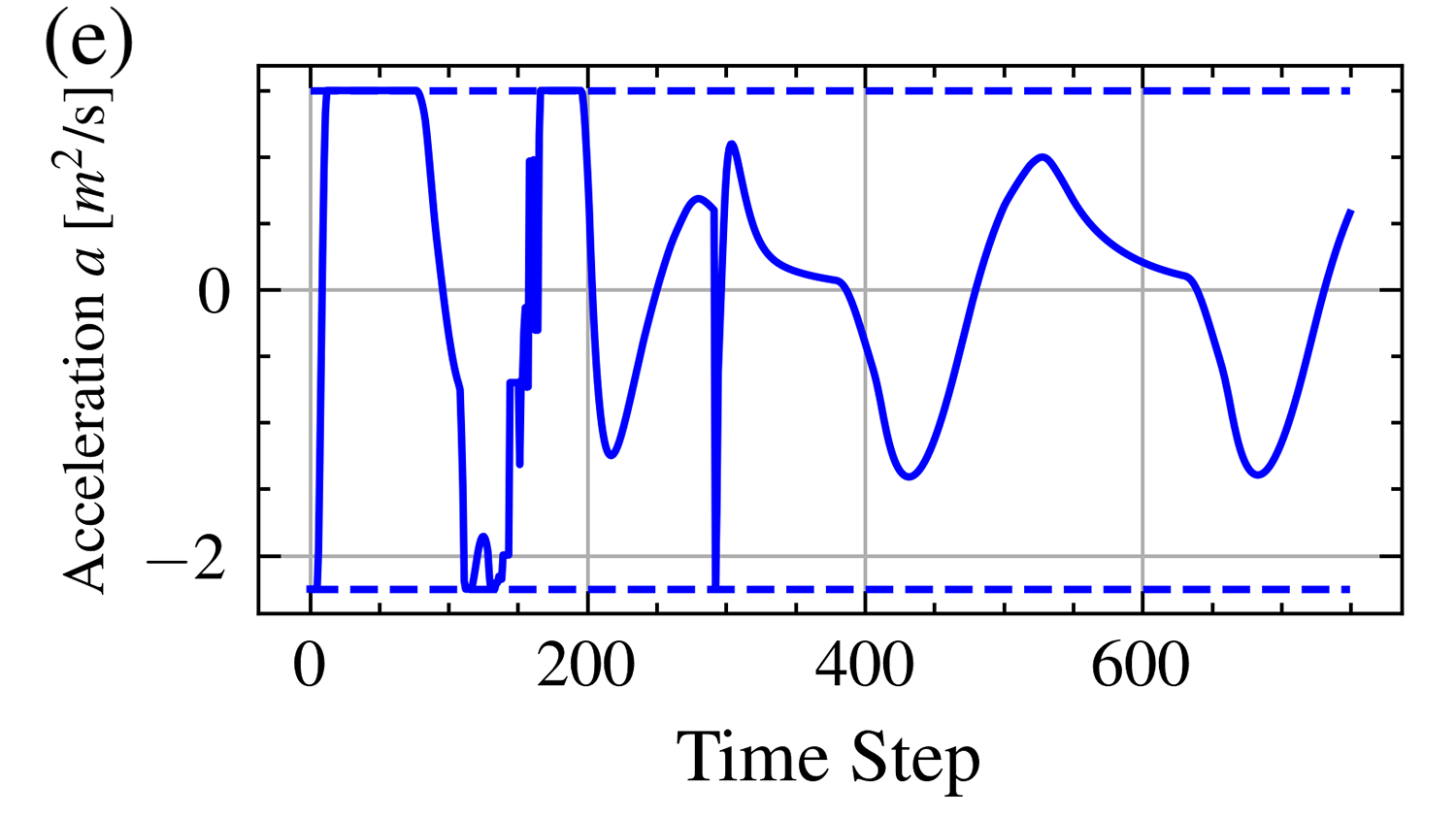}
        \label{fig:scn2_ctrl_acc}
    \end{subfigure}\vspace{-1.5mm}
    \hfill
    \begin{subfigure}[b]{0.49\textwidth}
        \centering
        \includegraphics[width=\textwidth]{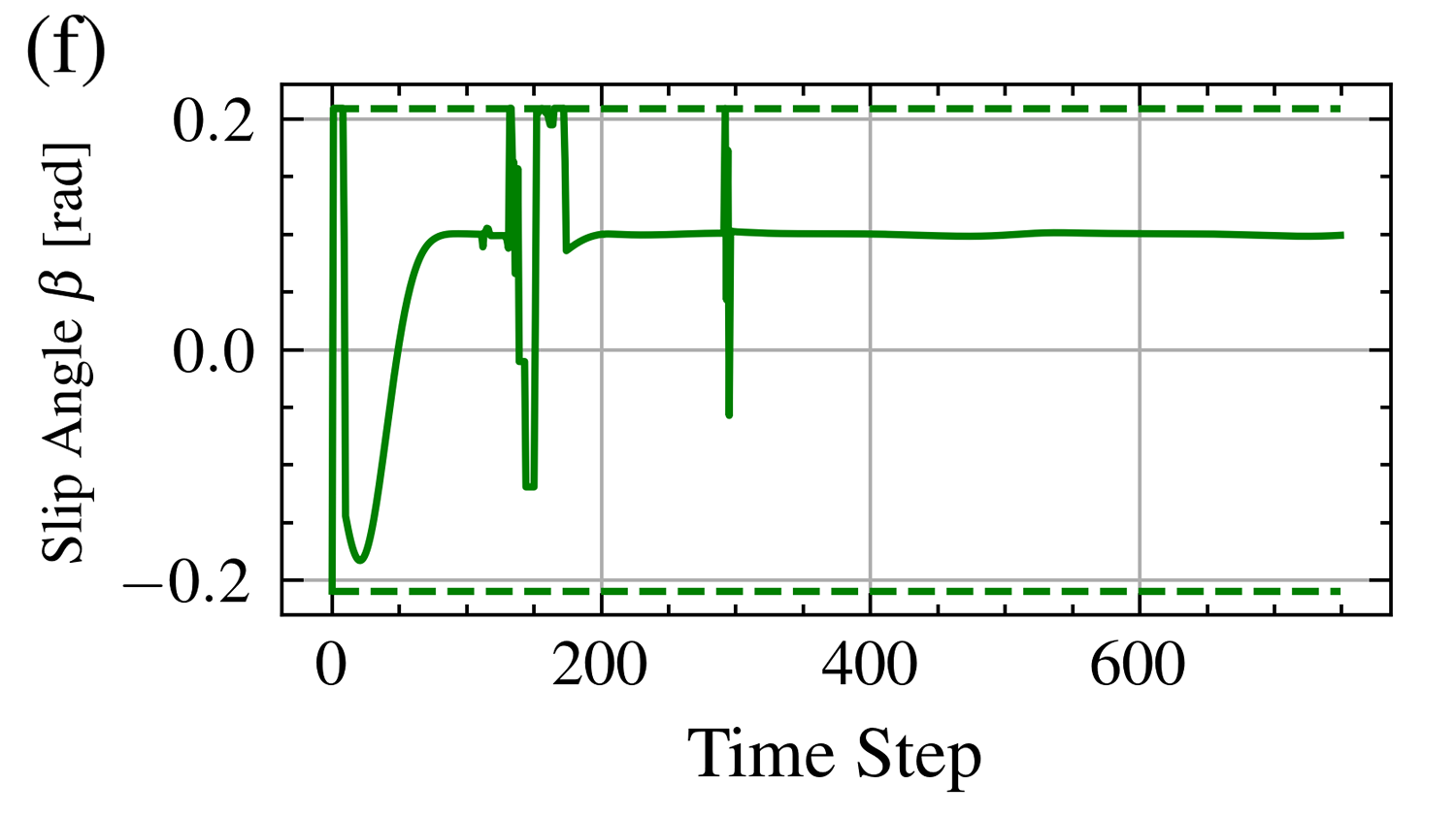}
        \label{fig:scn2_ctrl_omega}
    \end{subfigure} \vspace{-3.0mm}
    
    \begin{subfigure}[b]{0.50\textwidth}
        \centering
        \includegraphics[width=\textwidth]{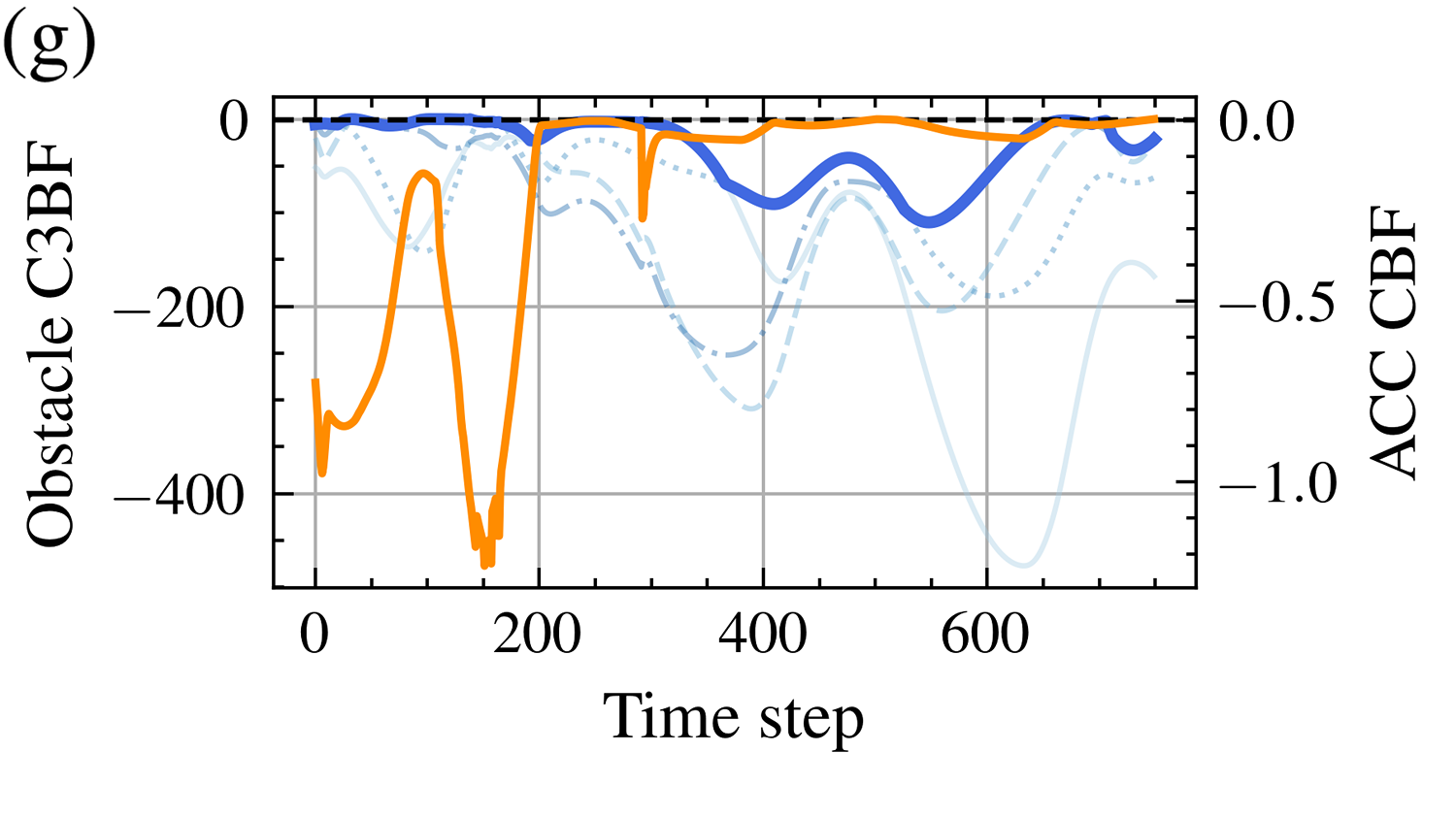}
        \label{fig:scn2_cbf}
    \end{subfigure} \vspace{-1.5mm}
    \hfill
    \begin{subfigure}[b]{0.48\textwidth}
        \centering
        \includegraphics[width=\textwidth]{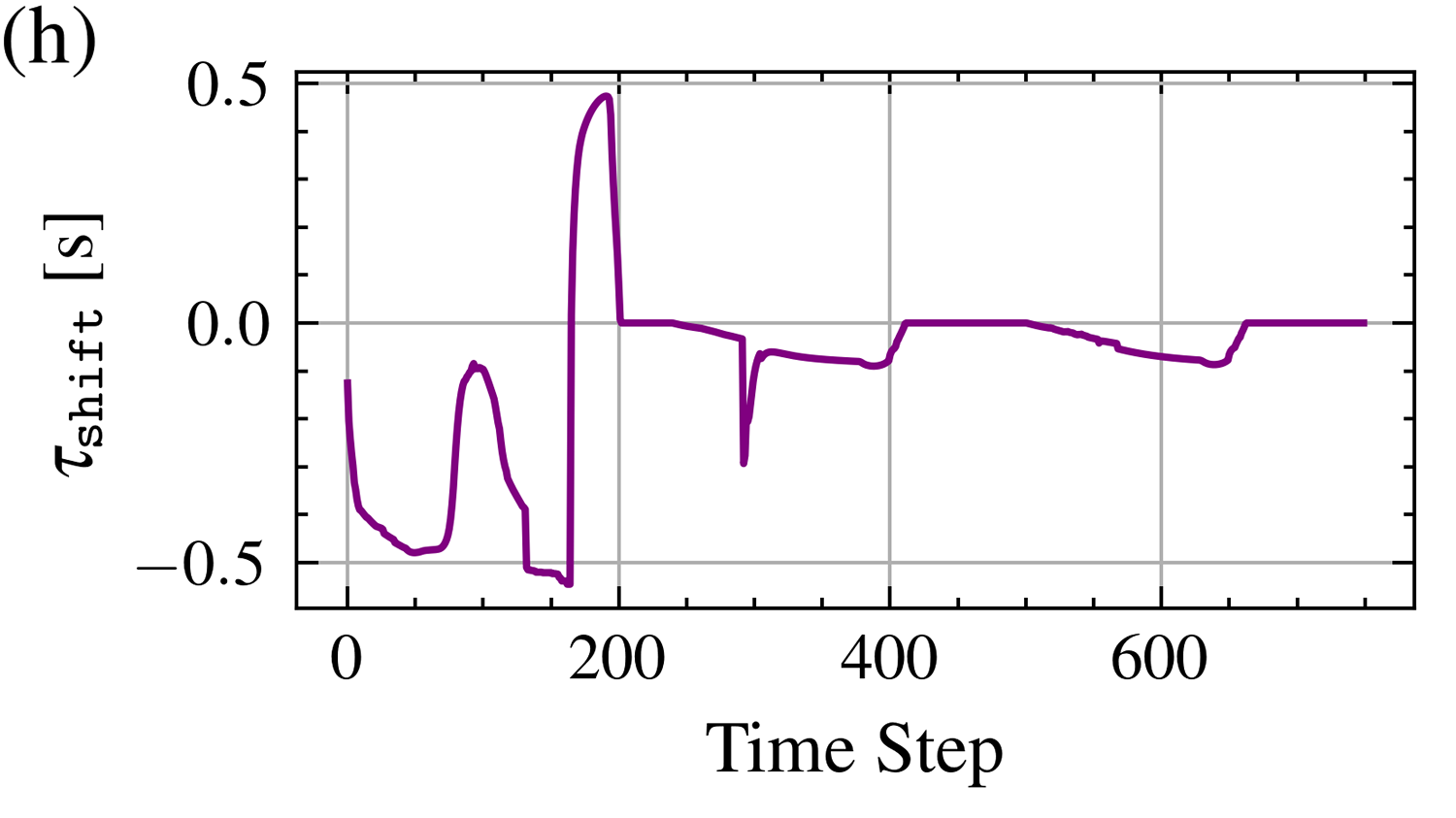}
        \label{fig:scn2_timeshift}
    \end{subfigure} \vspace{-6.0mm}
    \caption*{(B)} \label{fig:scen2}
\end{subfigure}

\caption{Representative trials (Panels A and B) selected from 50 simulation experiments. Each panel shows: (a) Baseline MPC-CBF trajectory, (b) TSG-guided MPC-CBF trajectory, (c) Relative position, (d) Relative velocity of the ego vehicle to the lead vehicle (red: baseline, blue dashed: TSG-guided), (e) Acceleration input history, (f) Slip angle input history, (g) CBF constraint values (orange: ACC, light blue: individual obstacles, blue: minimum value among obstacles), (h) Time shift parameter history.}
\label{fig:combined_scenarios}
\end{figure*}

%% file: _V.Experiments/a_sim_spec.tex
We implement the MPC-CBF framework~\eqref{eq:mpc-cbf-formulation} in Python using the IPOPT solver~\cite{fiedler_-mpc_2023}. The system dynamics~\eqref{eq:system_dynamics} are based on the kinematic bicycle model~\eqref{eq:kinematic_bicycle}, which assumes small slip angles (\( \sin\beta \approx \beta, \cos\beta \approx 1 \)). To enforce this assumption in practice, the slip angle \( \beta \) is constrained within a bounded range of \( \pm 12^\circ \). The maximum velocity constraint of $v_{\textup{max}} = 30~\textup{m/s}$ is applied, covering a wide range of urban and highway driving conditions. The control inputs are constrained to \(a_{\textup{min}} = -2.25 m/{s^2} \leq a \leq a_{\textup{max}} = 1.5 m/{s^2} \), representing typical acceleration and braking capabilities of passenger vehicles. The steering angle is limited to \( \pm 30^\circ \), which reflects the physical turning limits of the 2016 KIA Soul based on its wheelbase and turning radius~\cite{kiasoul2016specs}, as estimated via Ackermann geometry. The desired velocities for the ego and lead vehicles are set to 4.5 and 4.275 m/s, respectively, simulating low-speed driving scenarios such as stop-and-go traffic and parking lot driving. The reference trajectory is a circle of radius \( r_{\tt ref} = 15~\text{m} \), and the ego vehicle’s initial position is offset by approximately 10~m behind the lead. Each simulation involves 2 to 5 randomly generated obstacles, moving along radial directions with speeds uniformly sampled from \( [0.8, 1.4]~\text{m/s} \) interval while their radii are sampled from \( [0.3, 0.5]~\text{m} \).

The system is discretized with a time step of $\Delta t = 0.05~\textup{s}$, and the MPC controller uses a prediction horizon of 5 steps. The quadratic cost weight matrices are tuned to guide the ego vehicle along a nominal trajectory. To evaluate baseline performance, we verified that the ego vehicle could follow the nominal circular path and avoid collisions with the lead vehicle using a small CBF parameter, \( \gamma_{\tt{virtual}}\), set to 0.1. This configuration ensures the feasibility of the MPC-CBF problem, providing safe ACC under ideal conditions where the lead vehicle maintains a constant speed. Following~\cite{tayal2024collision}, the C3BF parameter used for obstacle avoidance is set to 1.0, \( \alpha(h) = \gamma h \), where \( \gamma = 1 \). A non-negative slack variable \( \delta_{\text{C3BF}} \in [0, 3.0] \) is introduced into the constraint, with a linear penalty term \( \lambda = 0.01 \) included in the cost to soften constraint violations when strict feasibility is difficult to maintain.

We note that our TSG-guided MPC-CBF is applied to the ego vehicle, which tracks a lead vehicle controlled by an MPC with a similar configuration. The lead vehicle controller uses the same dynamic model and objective as the ego, but replaces the CBF constraints with a Euclidean distance-based obstacle avoidance constraint. This ensures that the distance \( d_{\text{lead-obs}} \) between the lead vehicle and any obstacle remains greater than the sum of the lead vehicle's radius \( r_{\tt{lead}} \), the obstacle's radius \( r_{\text{obs}} \), and a safety margin \( d_{\tt{margin}} \).

\begin{table}[h]
\centering
\caption{Performance comparison over 50 Trials}
\label{tab:sim_results}
\begin{tabular}{lcc}
\toprule
\textbf{Controller} & \textbf{Success rate} & \textbf{Collision rate} \\
\midrule
Baseline MPC-CBF & 82\% & 18\% \\
TSG-guided MPC-CBF & \textbf{100\%} & 0\%  \\
\bottomrule
\end{tabular}
\end{table}

%% file: _V.Experiments/b_result.tex
We compare the performance of our proposed TSG-guided MPC-CBF approach, introduced in Section~\ref{subsec:optimization-problem}, with the baseline MPC-CBF approach in a scenario involving both a fluctuating-speed lead vehicle and multiple moving obstacles. This scenario incorporates both sudden lead vehicle maneuvers and interactions with moving obstacles.

In the considered scenario, the lead vehicle undergoes sinusoidal velocity variations, including sudden reversals and off-track deviations for obstacle avoidance, simulating highly unpredictable behaviors. The lead vehicle's velocity is capped at the desired speed. Meanwhile, multiple moving obstacles are initialized at random positions and constant-velocity trajectories are introduced into the environment, emulating the presence of pedestrians, cyclists, or other vehicles. The movement of the obstacles is visualized with color-changing trajectories to show their progression over time.

Figure~\ref{fig:combined_scenarios} illustrates two representative examples from this scenario, highlighting the ego vehicle's ability to maintain safety and trajectory tracking despite dynamic and uncertain interactions with both the lead vehicle and surrounding obstacles.

Fig.~\ref{fig:combined_scenarios}A illustrates that the baseline MPC-CBF (Fig.~\ref{fig:combined_scenarios}A(a)) is able to avoid dynamic obstacles but fails to anticipate a reversal in the lead vehicle’s trajectory to avoid obstacle collision, resulting in a rear-end collision. Safety constraint violations are observed in the inter-vehicle distance (Fig.~\ref{fig:combined_scenarios}A(c)). 
In contrast, the TSG-guided MPC-CBF (Fig.~\ref{fig:combined_scenarios}A(b)) dynamically adjusts the tracking reference using the time shift parameter (Fig.~\ref{fig:combined_scenarios}A(h)). Notably, changes in \(\tau_{\tt shift}\) occur in response to the lead vehicle's sudden reversal. This temporal adjustment modifies the ego vehicle's reaction, allowing it to maintain a safe distance without violating the constraint shown in Figs.~\ref{fig:combined_scenarios}A(e)--(g).

In Fig.~\ref{fig:combined_scenarios}B, the baseline controller (Fig.~\ref{fig:combined_scenarios}B(a)) reacts too late to a crossing obstacle, resulting in infeasible CBF conditions and a collision. Conversely, the TSG-guided controller (Fig.~\ref{fig:combined_scenarios}B(b)) proactively adapts to multiple lead vehicle reversals and dynamic obstacles, maintaining safety (Figs.~\ref{fig:combined_scenarios}B(e)--(g)). Adjustments in \(\tau_{\tt shift}\) (Fig.~\ref{fig:combined_scenarios}B(h)) are observed as the ego vehicle adjusts to the moving obstacles, demonstrating real-time responsiveness.

To evaluate robustness, we perform 50 randomized simulations for the above scenario, which involve a wide range of dynamic interactions between the ego vehicle and its environment. Table~\ref{tab:sim_results} summarizes the outcomes. The baseline MPC-CBF fails in 9 out of 50 trials, 6 due to collisions with the lead vehicle and 3 due to dynamic obstacle collisions. In contrast, the TSG-guided MPC-CBF completes all 50 trials without collision.




%% file: _VI.Conclusion/conclusion.tex


In this paper, we proposed a TSG-guided MPC-CBF controller for ACC aimed at non-straight-road scenarios. Simulation results demonstrate that the TSG-guided MPC-CBF safely navigated challenging scenarios, which incorporate a curved road track, time-varying constraints, and reversing lead vehicles, for which the traditional MPC-CBF violated constraints. Therefore, incorporating TSG, enforced safety constraints effectively in changing environments. The limitations in long-horizon scenarios and comparison study with other constraint control methods remain as future research.

